\documentclass[aip,amsmath,amssymb,reprint,]{revtex4-1}
\usepackage{xcolor}
\usepackage{graphicx}
\usepackage{hyperref}
\usepackage{float}

\draft

\newcommand{\PR}[1]{\textcolor{black}{#1}}

\newcommand{\rev}[1]{\textcolor{black}{#1}}
\newcommand{\revv}[1]{\textcolor{black}{#1}}

\newcommand{\kB}{k_\text{B}}
	\newcommand{\lD}{\lambda_\text{D}}
	\newcommand{\kD}{\kappa_\text{D}}
\newcommand{\lE}{\ell_E}
\newcommand{\lB}{\ell_\text{B}}
\usepackage{amsmath}

\newcommand{\vecb}[1]{\boldsymbol{\mathrm{#1}}}
\newcommand{\unitary}[1]{\boldsymbol{\mathrm{\hat #1}}}
\usepackage{amssymb}
\usepackage{amsbsy}

\newcommand{\di}{\mathrm{d}} 
\newcommand{\dpart}[2]{\frac{\partial #1}{\partial #2}}

\usepackage{comment}

\newcommand{\mean}[1]{\left\langle #1 \right\rangle}

\begin{document}
\title{Ionic association and Wien effect in 2D confined electrolytes}

\author{Damien Toquer}
\affiliation{Laboratoire de Physique de l'Ecole Normale Sup\'erieure, 24 rue Lhomond, 75005, Paris, France}
\author{Lyd\'eric Bocquet}
\email{lyderic.bocquet@ens.fr}
\affiliation{Laboratoire de Physique de l'Ecole Normale Sup\'erieure, 24 rue Lhomond, 75005, Paris, France}
\author{Paul Robin}
\email{paul.robin@ist.ac.at}
\affiliation{Institute of Science and Technology Austria, Am Campus 1., 3400 Klosterneuburg, Austria}

\date{4 October 2024}

\begin{abstract}
    Recent experimental advances in nanofluidics have allowed to explore ion transport across molecular-scale pores, in particular for iontronic applications.  {Two dimensional} nanochannels -- in which a single molecular layer of electrolyte is confined between solid walls --  {constitute a unique} platform to {investigate fluid and ion transport in extreme confinement, highlighting unconventional transport properties.}
        In this work, we study ionic association {in 2D nanochannels}, and its consequences on {non-linear ionic} transport, using {both} molecular dynamics simulations and analytical theory.  We show that under sufficient confinement, ions assemble into pairs or larger clusters in a process analogous to a Kosterlitz-Thouless transition, {here modified by the dielectric confinement}. We {further show that 
       the breaking of pairs results in an electric-field dependent conduction, a mechanism usually known as the second Wien effect. However the 2D nature of the system results in 
      non-universal,  temperature-dependent, scaling of the conductivity with electric field, leading to ionic coulomb blockade in some regimes.} 
        {A 2D generalization of the Onsager theory fully accounts for the non-linear transport.} 
        These results suggest ways to exploit electrostatic interactions between ions to build new nanofluidic devices.
\end{abstract}


\maketitle

\section*{Introduction}
Recent years have witnessed critical advances in the fabrication of atomic-scale fluidic channels\cite{bocquet_nanofluidics_2010,garaj_graphene_2010,lee_coherence_2010,feng_single-layer_2016,secchi_massive_2016,radha_molecular_2016,esfandiar_size_2017}, in an effort to achieve molecular control over water and ion transport -- similar to the transport machinery of cells. Considering the intense complexity of such nano-fabrication techniques, theoretical guidance has become necessary; however, traditional continuous models of fluid and charge transport breakdown under such extreme confinement\cite{bocquet_nanofluidics_2010,kavokine_fluids_2021,kavokine_ionic_2019,kavokine_interaction_2022,robin_nanofluidics_2023,richards_importance_2012,robin2023long,robin_modeling_2021}.

Particular attention has been drawn to 2D nanochannels recently\cite{radha_molecular_2016,emmerich_enhanced_2022} (Figure 1(a)). These systems, which are fabricated by van der Waals assembly, consist in flakes of a 2D material (graphene, MoS$_2$, hBN...) separated by graphene ribbons, creating an atomically-smooth and -thin channel permeable to water. Electrolytes confined inside such structures are expected to form a single ionic layer, which has been shown to enhance electrostatic interactions between ions\cite{kavokine_interaction_2022,robin_modeling_2021} and promote memory effects in conduction, such as the recently-demonstrated memristor effect\cite{robin2023long,robin_modeling_2021}.

The properties of ions in such systems have been studied using various techniques, both in theory -- mean field approaches\cite{robin_modeling_2021,levin_electrostatic_2002}, exact field theories\cite{minnhagen_two-dimensional_1987,robin_ion_2023}) --, and in simulations -- Brownian\cite{robin_modeling_2021, kavokine_interaction_2022}, molecular\cite{robin_modeling_2021,zhao_two-dimensional_2021} or \textit{ab initio} dynamics\cite{zhao_two-dimensional_2021}, and more recently machine learning forcefields trained with DFT-generated datasets\cite{fong_interplay_2024}. Overall, a growing body of evidence is shedding light on links between electrostatic correlations and non-linear ion transport \cite{avni_conductivity_2022}. A common blind spot of the aforementioned numerical approaches is the description of out of equilibrium ion transport. While theoretical predictions recently showed that 2D nanochannels should display non linear conduction due to electrostatic interactions between ions, numerical evidence for this has remained scarce, in particular due to the high computational cost of simulating large systems over long timescales. 

In this work, we implement molecular dynamics simulations to carry out a full description of ion association in 2D nanochannels, and show how electrostatic interactions impact charge transport at the molecular scale. We use a combination of Brownian dynamics (BD), where the solvent and channel walls are treated implcitly, and all-atom molecular dynamics (MD). The latter is particularly suited for identifying the impact of the discreteness of water molecules, while the former gives access to long-time dynamics by saving computational cost. We show that these two techniques offer complimentary tools to characterize ion association and ion-ion correlations in general. In both cases, we fully describe the formation of ionic clusters as function of the relative strength of electrostatic interactions compared to thermal noise. 

Simulations are complemented with an analytical framework of equilibrium and non-equilibrium properties, whose predictions are in excellent agreement with the numerical results.

This paper is organized as follows. In Section I, we describe our numerical implementation and how it can be used to quantify ionic pairing in confinement. 
Section II describes ionic association. We show how electrostatic correlations under confinement result in ionic pairing at thermal equilibrium, unveiling a transition between a fully paired state and a partially dissociated state as function of temperature, similar to a modified Kosterlitz-Thouless transition.
In section III, we explore out-of-equilibrium properties and demonstrate non-linear transport of ions in the monolayer, governed by pair breaking under an electric field (a phenomenon known as the second Wien effect). 

\section{Methods}

\rev{In this work, we model an electrolyte confined in a 2D nanochannel of height $h$, made of carbon walls (Figure 1a). We assume that the system is connected to macroscopic fluidic reservoirs where Ag/AgCl electrodes are located, imposing a voltage drop across the system and creating a static electric field $\vec E$.}

\subsection{MD and BD dynamics}

\begin{figure*}
    \centering
    \includegraphics{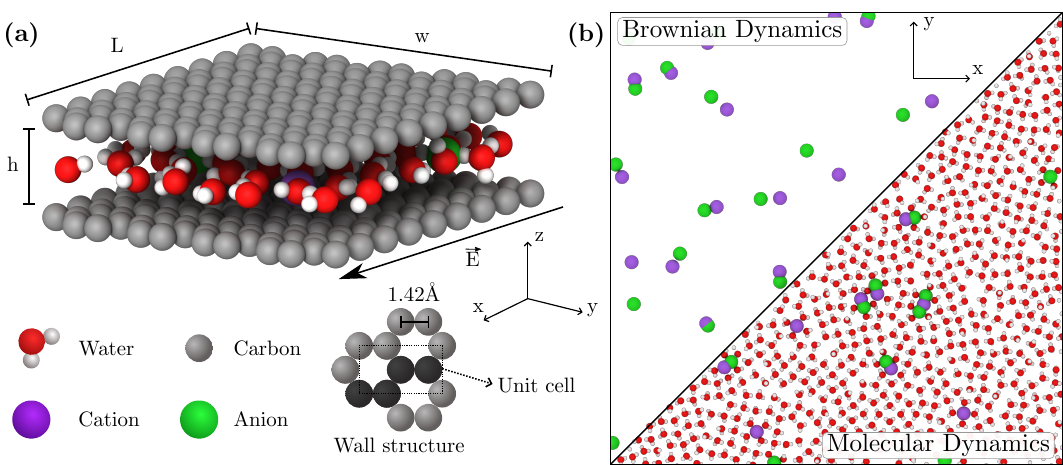}
    \caption{\label{fig:scheme}System studied in this paper. (a) Simulation setup for the molecular dynamics simulation. The wall are generated from a unit cell of 4 carbon atoms, in black in the structure. (b) Comparison of Brownian dynamics (top left) and molecular dynamics (bottom right) simulations at a same ionic density.}
\end{figure*}

Two different simulation methods are used throughout this paper. In all-atom molecular dynamics (MD), the entire system (water molecules, ions and atoms from the channel walls, see Figure 1(a)) is simulated using classical Lennard-Jones and Coulombic forcefields and solving Newton's equations of motion. In Brownian dynamics (BD), on the other hand, water and channel walls are treated implicitly, and only the positions of ions are tracked using overdamped Langevin equations (Figure 1b). \rev{This is motivated by the finding that electrostatic interactions on ions from the walls and the solvent can be accounted for in a coarse-grained way by renormalizing the ion-ion interaction potential \cite{kavokine_interaction_2022}. Further details regarding the simulation methodology can be found in Supplementary Information (SI).}

\rev{
\subsection{Interaction potentials}
In MD dynamics, as stated above, all particles (ions, wall atoms and water molecules) interact both through electrostatic and Lennard-Jones interactions, the latter representing attractive and repulsive van der Waals forces. The interaction energy between particles $i$ and $j$ thus reads:
\begin{equation}
	V_{ij}(r) = 4 \epsilon \left[\left(\frac \sigma r \right)^{12} - \left(\frac \sigma r \right)^{6} \right] + \frac{q_i q_j}{4 \pi \epsilon_0 r},
\end{equation}
where $q_i$ is the charge of particle $i$, $r$ the interparticle distance and $\epsilon$ and $\sigma$ some particle-specific constants detailed in SI.}

\rev{
One of the main shortcomings of MD dynamics is that only a fraction of the simulation time is spent on simulating ion dynamics -- the focus of this work. Integrating out solvent degrees of freedom allows ton run much faster implicit-solvent BD simulations. This is usually done by modelling water as a dielectric medium of constant $\epsilon_w = 80$; however, here one must accounts for atomic confinement as well. Walls not only restrict ion transport geometrically, but also modify the electric potentials generated by ions due to having a much lower permittivity compared to water (in what follows, we will assume that the walls have a dielectric constant $\epsilon_s = 2$) -- this process, called interaction confinement \cite{kavokine_interaction_2022}, is caused by the fact electric field lines from ions cannot fully penetrate the walls and instead concentrate within the fluid. Overall, this results in a renormalized ion-ion potential given by\cite{kavokine_interaction_2022}:
\begin{equation}
	V_{ij}(r) = \frac{q_i q_j}{4 \pi \epsilon_0 \epsilon_w r} \int_{0}^{\infty} J_0(u) \frac{\tanh \left[ u \frac h {2 r}\right] + \frac{\epsilon_w}{\epsilon_s}}{ \frac{\epsilon_w}{\epsilon_s}\tanh \left[ u \frac h {2 r}\right]  + 1},
	\label{eqn:MD}
\end{equation}
where $J_0$ is a Bessel function and we recall that $h$ is the height of the channel. For $\epsilon_w / \epsilon_s \gg 1$ and $r \gg h$, the above integrand can be cut into $0 < u < A$ and $u >A$ (defined by $\int_0^A J_0(u) \, \di u = 0$, $A \simeq 1.1$). Assuming that the latter part oscillates quickly and integrate to zero, and that $\int_0^A J_0(u) \times [...] \simeq \int_0^A \frac 1 A \times [...]$, we obtain:
\begin{equation}
	\beta V_{ij}(r) = -\frac{\text{sign}(q_iq_j)}{T^*}\log\left(\frac{r}{r +\xi}\right),
	\label{eqn:BD}
\end{equation}}
where we introduced the inverse temperature $\beta$ as well as two other parameters: the dielectric length:
\begin{equation}
    \xi = A \epsilon_w h/2 \epsilon_s,
\end{equation} and the (dimensionless) reduced temperature $T^*$ defined by:
\begin{equation}
	T^* = \frac{2\pi\varepsilon_0\varepsilon_wh}{Z^2e^2 A}k_BT,
\end{equation}
with $Z$ the valency of ions and $e$ the elementary charge. This parameter plays the role of a coupling constant; we will use it throughout the manuscript as a measure of the strength of electrostatic interactions. One finds that $T^* \simeq 0.4$ for monovalent ions at room temperature and $T^* \simeq 0.1$ for divalent ions. In practice, we use this potential in BD simulations, with $\xi \simeq 14\,$nm. It should be noted that, in the limit $r \ll \xi$, one recovers $V_{ij}(r) \simeq q_i q_j/4 \pi \epsilon_0 \epsilon_s r$ as field lines now fully permeate the solid.

When comparing with MD simulations, one should notice that the system is confined between two vacuum slabs (as walls are made of a single layer of carbon atoms), corresponding to $\epsilon_s = 1$ and $\xi \simeq 28\,$nm. Throughout the paper, to simplify discussion, we always assume that $Z=2$ and we instead use temperature as a tuning parameter of the interactions.

\rev{
Confinement has been shown to reduce the effective permittivity of water or make it anisotropic\cite{fumagalli_anomalously_2018,schlaich_water_2016,monet2021nonlocal}. The exact dielectric properties of confined water are still debated however, with some recent experiments showing an increase in in-plane permittivity instead\cite{wang2024plane}. It has been shown elsewhere \cite{robin_modeling_2021} that dielectric anisotropy only has a weak effect on ion-ion interactions. Besides, our results can be extended in a straightforward way to other values of $\epsilon_w$ in confinement by tuning $T^*$.
}

\rev{
Lastly, while in this paper we only consider the case of non-conducting walls made of a dielectric material, the impact of wall conductivity has been discussed elsewhere\cite{kavokine_interaction_2022,rochester2013interionic,kondrat2011superionic}. Briefly, conduction electrons within the walls are able to screen off ion-ion electrostatic interactions, reducing their range. 
}

\subsection{Ionic current}
For a given value of the imposed electric field $\vec E$, we extract the ionic current density from simulated trajectories through:
\begin{equation}\label{eq:curextraction}
    j = Ze\frac{N}{h\mathcal{A}}\left(\langle v_+\rangle-\langle v_-\rangle\right),
\end{equation}
where $\mean{v_+}$ and $\mean{v_-}$ are respectively the time-averaged velocity of the cations and the anions along the direction of the field, $N$ is the number of ions and $\mathcal A$ the 2D area spanned by the channel. We can deduce the conductivity through: 
	\begin{equation}
    \sigma(E) = \frac{j(E)}{E}.
\end{equation}
Which we will compare to the Ohmic conductance (defined for ions behaving like ideal tracer particles):
\begin{equation}
    \sigma_0 = \frac{2N}{h\mathcal{A}}\frac{(Ze)^2D}{k_BT},
\end{equation}
with $D$ the diffusion coefficient of ions, assumed to be the same for anions and cations (\rev{in the case of a difference in diffusivity, $D$ should be replaced by the average of the coefficients of cations and anions}). Parameters of BD simulations are chosen so that $D/\kB T$ is independent of temperature, so that we can use the same value of $\sigma_0$ to compare the conductivity of the system for different strengths of electrostatic interactions.

\subsection{Ionic pairs and clusters}

We observe in BD simulations that at low temperature, ions tend to form pairs. Since they are neutral and do not contribute to conduction, we use conductivity measurements as a way of quantifying the fraction of free ions $n_f$:
\begin{equation}
	n_f = \frac{\sigma(E)}{\sigma_0}.
	\label{eqn:nf}
\end{equation}
This definition of ionic pairing has the advantage of being the closest to what could be measured in experiments, in addition to being very convenient to compute. In fact, it effectively counts the fraction of charge carriers only if we can neglect higher-order correlations between ions (which tend to impede conduction compared to the Ohmic case), or equivalently if we assume the mobility of individual ions not to vary with the field. We make this assumption in the following, since we expect the non linearity of the system to be dominated by the variation of the fraction of charge carriers (second Wien effect), and not by the variation of the mobility (first Wien effect\cite{kaiser_wien_2014,kaiser_onsagers_2013}, and Debye-Hückel correlations in general\cite{berthoumieux_non-monotonic_2024}). This intuitive definition unfortunately fails if $E$ is too low, due to a poor signal-to-noise ratio. \rev{In SI, we report an alternative definition based on correlation functions, which is consistent with Eq. \eqref{eqn:nf} but can be extended without problem to $E =0$. }

In MD simulations, we observe the formation of large ionic clusters, in addition to ionic pairing. \PR{Some of these clusters bear a non-zero charge, and therefore participate to conduction. We therefore distinguish between two quantities: first, the fraction $n_f$ of free ions; and the fraction of charged clusters $n_f^\text{cluster}$ (which includes free ions -- as they are charged clusters of size 1).} In practice, we find that all clusters containing an even number of ions are electrically neutral, and odd-sized clusters have a charge $\pm 1$ (as clusters with multiple charge defects are highly unstable). From there, we can deduce that a measure of the fraction of charge carriers density is:
\begin{equation}
	n_f^{\text{cluster}} = \frac{\sum_{l\text{ odd}}N_l}{2N},
\end{equation}
with $N_l$ the number of clusters of size $l$, which we determine directly from trajectories through network analysis. \PR{The fraction of free ions is defined as:
\begin{equation}
	n_f = \frac{N_1}{2N}.
\end{equation}
This exact procedure used to compute $N_l$ from trajectories is described in the Supplementary Information.}

\section{Pairing transition and electrostatic screening}\label{sec:1}
\rev{In this section, we quantify the formation of ionic pairs using both simulations and theoretical analysis. We start by predicting that the system undergoes a conductor/insulator phase transition reminiscent of the 2D Kosterlitz-Thouless transition. We analytically predict the system's critical temperature, which is found to {slightly} deviate from the usual KT result due to the quasi-2D nature of confined electrolytes. We then compare those results to simulations. We find that indeed at low temperature, almost all ions assemble into pairs. We characterize this transition numerically, and show it is of infinite order as predicted by the KT analogy.}

\subsection{Theoretical analysis}

\subsubsection{Critical temperature and correlation length}
\rev{It has been shown\cite{levin_electrostatic_2002} that the 2D Coulomb gas, made of charged particles interacting through a potential $V(r) = - \frac{1}{T^*}\log r$, undergoes a pairing phase transition for $T^* < 0.25$ that is equivalent to the Kosterlitez-Thouless transition. Here, we sketch the outline of such a derivation, while focusing on differences between the exact 2D Coulomb gas and our actual system -- 2D confined electrolytes.}

At the mean-field level, we can treat ion pairs as a separate, ideal chemical species, and write down the chemical equilibrium with free ions:
\begin{equation}
	n_p = 1 - n_f = \zeta n_f^2 e^{2\beta\mu_e}.
	\label{eqn:ChemEqu}
\end{equation}
Here, $n_p$ is the fraction of ions that are part of a pair, $n_f$ is the free ion fraction, $\mu_e$ is the excess chemical potential of free ions, $\beta=1/k_BT$ and $\zeta$ is the internal partition function of a pair. In the 2D Coulomb gas, $\mu_e$ can be obtained through a Debye charging process, by computing the field created on a given ion by its surrounding Debye atmosphere:
\begin{equation}
	2\beta\mu_e = \frac{1}{T^*} \left[ \frac{K_0(\kD r_0)}{\kD r_0K_1(\kD r_0)}\right],
	\label{eqn:Psi}
\end{equation}
where $r_0$ is a measure of the ionic size\rev{, $K_0$ and $K_1$ are modified Bessel functions of the second kind} and
\begin{equation}
	\kD = \sqrt{\frac{4 \pi n_f c}{T^*}},
\end{equation}
is the inverse Debye length at salt concentration $c$ (expressed in atoms per surface area). For $T^* < T^*_c = 0.25$, Eq. \eqref{eqn:ChemEqu} becomes $1 - n_f \propto n_f^{2 - 1/2T^*}$, with $2 - 1/2T^* <0$, and therefore admits no solution. This is the signature of the pairing transition. In addition, just above the transition, one can compute the correlation length through the Debye length (in the limit where the salt concentration is not too high \cite{rotenberg_underscreening_2018}):
\begin{equation}
	\lD = \sqrt{\frac{T^*}{4 \pi n_f c}} \underset{T^* \to T^*_c} \propto e^{\frac{T^*/4}{T^* - T^*_c}}.
    \label{eq:debyescaling}
\end{equation}
The correlation length therefore depends in a non-algebraic manner on temperature, and in particular no critical exponent can be defined -- this shows that the phase transition is of infinite order. \rev{Interestingly, the system possesses a critical \textit{line}, meaning that there is a single value of the phase transition temperature regardless of salt concentration. This line ends at a given value of concentration, depending on the precise value of $\zeta$ and the exact criterion used to define ion pairs (for our choice of parameters, the transition occurs independently of concentration for $c < 0.9\,$M).}

\rev{
In our case, however, ionic interactions are not exactly logarithmic: $V(r) \sim - \frac 1 {T^*}\log r/(r + \xi)$, see Eq. \eqref{eqn:BD}. Repeating the above argument, we find that:
\begin{equation}
	2\beta\mu_e = \frac{1}{T^*} \left[\frac{\xi}{ \kD r_0 (r_0 + \xi)} \frac{K_0(\kD r_0) - K_0(\kD (r_0 + \xi))}{K_1(\kD r_0) - K_1(\kD (r_0 + \xi))}\right].
	\label{eqn:PsiXi}
\end{equation}
Then, Eq. \eqref{eqn:ChemEqu} stops having a solution for
\begin{equation}
	T^*_c = -\frac 1 4 \frac{\xi }{\xi + 1}\min_k k \dpart{g_\xi}{k}
\end{equation}
where we defined 
\begin{equation}
	g_\xi(k) =  \frac{K_0(k) - K_0(k (1 + \xi/r_0))}{K_1(k) - K_1(k (1 + \xi/r_0))}
\end{equation}
We notably find that $T^*_c \simeq 0.18$ for $\xi/r_0 = 14$ and we recover $T^*_c = 1/4$ for $\xi \to \infty$.
}

\rev{Lastly, let us discuss the case where the dielectric length $\xi$ is short, such as for $\epsilon_s \sim \epsilon_w$. In that case, interactions are weakened as the electric field lines of ions leak into the walls. This situation is similar to confinement by conducting or metallic walls \cite{kavokine_interaction_2022,rochester2013interionic,kondrat2011superionic}. While some degree of ionic pairing was reported in such context, it is a priori much weaker than for channels with high dielectric contrast such as here. }
\rev{
\subsubsection{Discussion and scaling laws}
In the previous paragraph we discussed the formation of ion pairs in confined electrolytes. These pairs, called Bjerrum pairs, are usually absent from bulk aqueous electrolytes; we thus discuss the impact of confinement below.}

In any setting, the strength of the ionic interactions can be quantified by the Bjerrum length\cite{bjerrum_untersuchungen_1926} $l_B$, defined as:
\begin{equation}
	\beta V(\lB) = \kB T.
\end{equation}
It is the typical length at which electrostatic interactions become comparable to thermal agitation. For a bulk system, the Bjerrum length is then given by:
\begin{equation}
	\lB^\text{bulk} = \frac{e^2}{4\pi\epsilon_0\epsilon_wk_BT},
\end{equation} 
while for a 2D confined electrolyte, where electrostatic interactions are quasi-logarithmic (see Eq. \eqref{eqn:BD}), it reads:\rev{
\begin{equation}
		\lB^{\text{2D}} = \frac{\xi}{e^{T^*}-1} {\sim}\frac{\xi}{T^*}.
\end{equation}
}
This can be used to compare the interaction in each systems:
\begin{equation}
	\frac{\lB^{\text{2D}}}{\lB^{\text{bulk}}} \simeq \frac{\varepsilon_w}{\varepsilon_s} \gg 1.
\end{equation}
This notably shows how, under confinement, dielectric contrast between the walls and water greatly reinforces electostatic interactions. In bulk water, $\lB^{\text{bulk}} = 0.7$nm, meaning that long-range interactions between ions are essentially negligible compared to thermal noise. In the slit this is not the case anymore, as the Bjerrum length is orders of magnitude larger: for our choice of parameters, we obtain $\lB = 14\,$nm, far exceeding any other microscopic lengthscale. As we will discuss for several later points, 2D confined electrostatic interactions are nearly scale-invariant, owing to their quasi-logarithmic nature.

\rev{This fact allows us to understand the emergence of the phase transition using scaling laws. The excess chemical potential $\mu_e$ of free ions can be thought of as half the free energy cost of breaking a pair. Since the free cation and anion obtained by breaking a pair are on average distant by $\lD$, we can assume that:
\begin{equation}
	2 \beta \mu_e \sim \frac{1}{T^*}\log \lD/r_0,
	\label{eqn:FreeEnergyCost}
\end{equation}
where $r_0$ is the ionic size. At low enough temperature, we expect that most ions are paired, so that $n_p = 1 - n_f \simeq 1$. From the chemical equilibrium \eqref{eqn:ChemEqu}, and using the definition of $\lD$, we obtain:
\begin{equation}
	n_f \propto c^{\frac{1}{4T^* - 1}}.
\end{equation}
This leads to a contradiction if $T^* < 0.25$: for low enough salt concentration $c$, $n_f >1$, which is impossible -- the chemical equilibrium is broken and there is a phase transition.}

\subsection{Comparison with simulations}
\subsubsection{Brownian dynamics simulations}

\begin{figure*}
    \centering
    \includegraphics{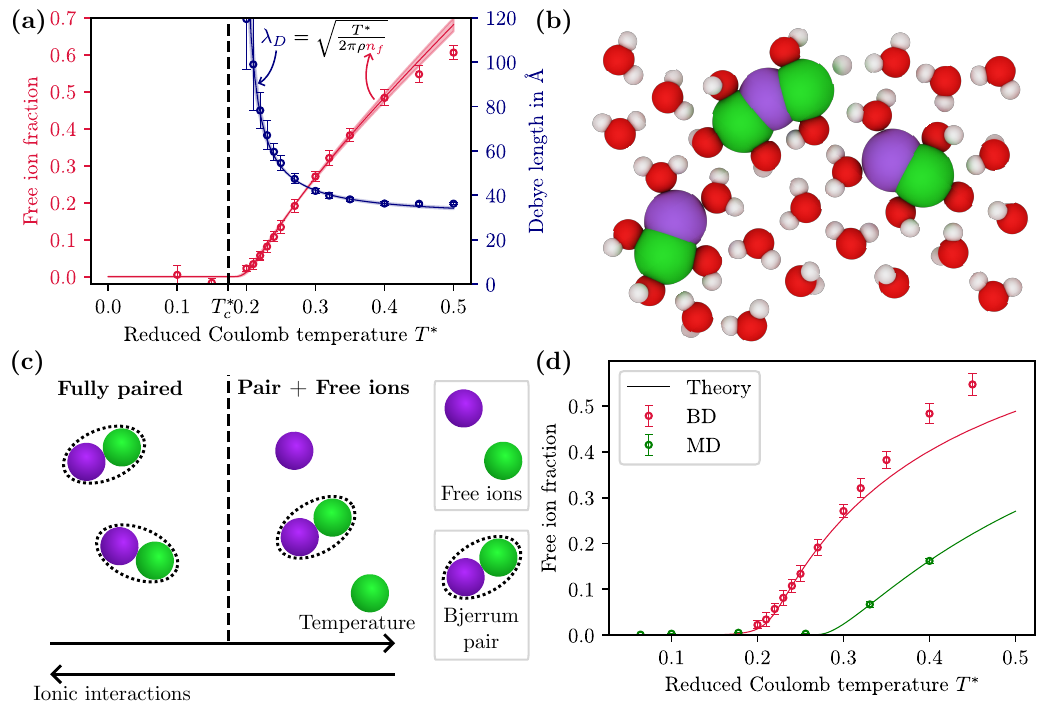}
    \caption{\label{fig:transition}Pairing transition in theory and simulations. (a) (Red points) Free ion fraction from BD simulation for various reduced Coulomb temperature. (Red line) Fit of the free ion fraction for an infinite order phase transition. (Blue points and line) Debye length obtained from the the free ion fraction using Equation \eqref{eq:debyescaling}. (b) Screenshot from MD simulations at equilibrium for $T^* = 0.1$. We observe large ionic cluster (7 ions in the screenshot) that move together and than can have a net charge (-1 in the screenshot). (c) Schematic of the phase transition. At small reduced Coulomb temperature or equivalently high ionic interaction, the system is fully paired. Otherwise, there is some remaining free ions. (d) (Red points) Free ion fraction from BD simulations for various reduced Coulomb temperature. (Red line) Theoretical curve for $r_0 = 1.2$ nm and $\zeta = 0.22$. (Green points) Free charge carrier density from MD simulations for various reduced Coulomb temperature. (Green line) Theoretical curve for $r_0 = 0.18$ nm and $\zeta = 0.33$.}
\end{figure*}

In order to verify our theoretical results with our BD simulations, we plot in Figure \ref{fig:transition}.(a) the evolution of the free ion fraction with the reduced Coulomb temperature for $\xi = 13.8$\ nm. As we lower the temperature, the free ion fraction decrease, up to \PR{an almost fully paired system for a finite value of temperature. Quantitatively, one can define the crtitical temperature in BD simulations as the temperature for which the system's correlation length diverges. In Fig. \ref{fig:transition}.(a), we show the evolution of the Debye length $\lD$ with temperature, and we fit data from BD simulations with an ansatz inspired by Equation \eqref{eq:debyescaling} (solid blue line):
\begin{equation}
	\lD(T^*) \propto \exp \left[\frac{T^*/4}{T^* - T^*_c}\right]
\end{equation}	
We find $T_c^* = 0.17$, which is very close to the theoretical value $T^*_c = 0.18$. The comparison between theory and simulations notably confirms that the correlation length does diverges in a non-algebraic way close to the transition, validating the fact it is of infinite order.
}
\PR{The comparison between the fraction of free ions in simulations and theory (see Equation \eqref{eqn:ChemEqu}) shows similarly good agreement, as displayed on Figure \ref{fig:transition}.(d) (red symbols and solid line).
}

\subsubsection{Molecular dynamics simulations and effect of the short-range interactions}
In previous sections, \PR{we discussed BD simulations where ions do not have any repulsion potential, meaning that short-distance structures could be adequately resolved}. Without repulsive potential, the distance between ions in a pair fluctuates around 0. As there is no permanent dipole in this case, their interactions with other ions are very weak. In all atom MD, where ions also interact through repulsive LJ interactions, pairs fluctuates around a finite distance (see Figure \ref{fig:scheme}.(b) and Figure \ref{fig:transition}.(b)). In this case, interaction with pairs are no longer negligible, allowing the formation of more complex structures.

Other works\cite{robin_modeling_2021,zhao_two-dimensional_2021} that used different simulations techniques reported various possible structures of ionic clusters, depending for example on the chemical nature of ions\cite{robin_modeling_2021, zhao_two-dimensional_2021} or of the wall\cite{fong_interplay_2024}. \rev{In our case, ion correlation functions reveal a short-distance structuration due to contact and solvent-separated ion binding, up to clusters of size $\sim 6$ in absence of electric fields (Figure S3).} Overall, the relations between short-range interaction and short-range structure seem non universal and is in practice difficult to analyse. Instead, we can focus on the free ion fraction, in comparison with BD simulations.

\rev{We find that, like in BD, the free ion fraction $n_f$ drops to zero below a critical temperature, as shown in Figure \ref{fig:transition}.(d) (green symbols). The behavior of $n_f$ with temperature is well-described by theoretical predictions; however the value of the critical temperature is found to differ from BD simulations, with $T^*_c \simeq 0.24$. This discrepancy can be attributed to two facts. First, in MD simulations, we only account for a single layer of wall atoms: the electrolyte is thus effectively confined by a slab of empty space, with $\epsilon_r = 1$, leading to a much higher value of the dielectric length $\xi = 28\,$nm. Besides, the minimal inter-ion approach distance is now unequivocally set by the physical size of ions, set by Lennard-Jones short-distance interactions.  For our choice of simulation parameters, the distance of physical contact is $r_0 = 0.18\,$nm (corresponding to approximately twice the ionic radius of sodium). For these values of $\xi$ and $r_0$, we obtain from our theoretical model $T^*_c = 0.24$, very close to the value infered from simulations (Figure \ref{fig:transition}.(d), green line). }
	
	\rev{
To obtain the theoretical curves of Figure \ref{fig:transition}.(d), one must also specify the quantity denoted $\zeta$ in Eq. \eqref{eqn:ChemEqu}. It can be interpreted as a ``volume fraction'': if one fixes the position of the first ion (assumed to be a disk of radius $r_0/2$), one must place the second ion exactly at a distance $r_0$ from the first ion's center to form a contact pair. Within that disk of radius $r_0$, the pair occupies a fraction $\zeta_\text{MD} \simeq 0.36$ (corresponding to the central ion and around half of the second one). If instead ions are allowed to overlap, as in BD, then that fraction drops to approximatively $\zeta_\text{BD} \simeq 0.25$ as paired ions almost fully overlap. In practice, we treat $\zeta$ as a fitting parameter for the sake of simplicity, and obtain $\zeta_\text{MD} = 0.33$ and $\zeta_\text{BD} = 0.22$, very close to estimated values. 
}

\rev{Similarly, we show in SI that the fraction of charged clusters also exhibit a drop below $T^*_c$. However, we find that the equilibration time of clusters grows rapidly as temperature is lowered: due to our limited simulation time, we find that $n_f^\text{cluster}$ is not exactly zero at very low temperature.
}

\rev{It should be noted that in our theory, the chemical equilibrium (Eq. \eqref{eqn:ChemEqu}) should in principle be updated to take the formation of clusters into account, instead of only ion pairs. In practice, however, large cluster only represent a small fraction of ions (see Figure S3) and Eq. \eqref{eqn:ChemEqu} can be used as an approximation. }

\section{Non linear transport}
In this section, we {study the out-of-equilibrium properties of the 2D confined systems under electric drivings.}
In particular, we focus on the increase in conductivity caused by the field-induced dissociation of ions pairs, a process known as the second Wien effect. We show analytically that this phenomenon is at the source of the non-linear ion transport, as the ionic current across the nanofluidic slit becomes a power law of the applied voltage, with a non-universal exponent, {different from the 3D case}. Brownian dynamics validate these theoretical predictions and notably the strong dependence of the exponent with the reduced Coulomb temperature.

\begin{figure*}
    \centering
    \includegraphics{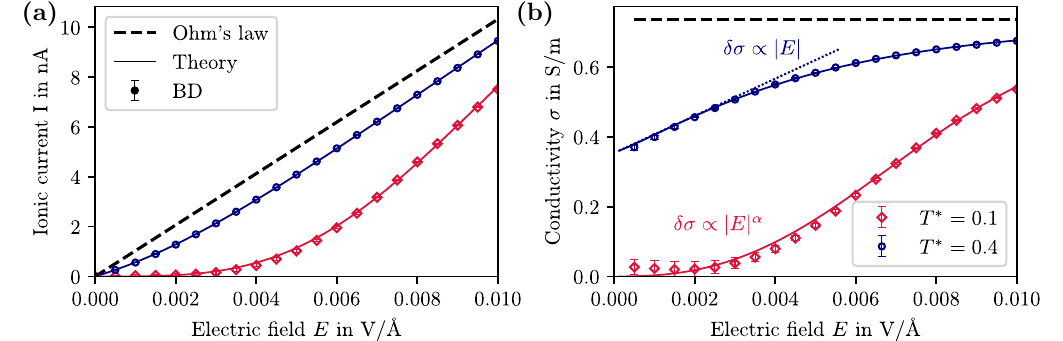}
    \caption{\label{fig:transport_curve} Comparison between BD transport simulations and theory. (a-b) Ionic current (a) and conductivity (b) for reduced Coulomb temperature of 0.1 (red) and 0.4 (blue), compared with Ohm's law (dashed black line). Symbols: BD simulations. Solid lines: Theoretical model, Equation \eqref{eqn:ClusterizedNf}.}
\end{figure*}

\rev{
\subsection{Non-linear behavior of the ionic current}
As previously stated, the application of an external {electric} field across the 2D slit tends to break ion pairs, effectively increasing the electrolyte's conductivity. Numerically, we can quantify this non-linear contribution by subtracting from the total ionic current, the Ohmic term resulting from the conduction of ions that are free in absence of field:
\begin{equation}
    \delta j(E) \simeq j(E)-n_f(0)j_0 = \delta n_f(E)j_0,
\end{equation}
with $j_0 = \sigma_0 E$ the current density of a fully dissociated electrolyte. {In terms of conductivity, this yields to the following contributions:
	\begin{equation}
    \sigma(E) = \sigma_0 + \delta\sigma(E)
\end{equation}}
Figure \ref{fig:transport_curve}.a shows the conductivity as a function of the electric field below and above the transition. When $E\to 0$, we observe that the conductivity tends linearly to a constant above the transition $T^*>T_c^*$, and vanishes like a power law below the transition. \rev{In this regime, we will define the exponent $\alpha(T^*)$:
\begin{equation}
    |\delta j(E)| \underset{E\to 0}{\propto} { |E|^{\alpha(T^*)}}.
\end{equation}} In the limit of large applied fields, we instead observe that $n_f\to 1$: we recover Ohm-like conduction, as in fully dissociated electrolytes. The corresponding ionic current $I$ is plotted in Figure \ref{fig:exponent}.b.}

\subsection{Theory}

In this section, we derive analytically the expression of the non-universal exponent $\alpha(T^*)$. Our approach is based on {an extension of} Onsager's study of the second Wien effect for bulk {3D} weak electrolytes \cite{onsager_deviations_1934}. {The generalization to 2D has been first developed} in Ref. \cite{robin_modeling_2021} {and we elaborate on this description here to compare with our numerical results}. 
In what follows, we restrict ourselves to the case of the ideal 2D Coulomb gas ($\xi = \infty$), and {focus on the paired regime}, ($T^* < T^*_c$).

\subsubsection{An Onsager's approach of the 2D Wien effect}
We assume that the fraction $n_f$ follows a generic evolution equation, introducing a dissociation (resp. association) \rev{time} $\tau_d$ (resp. $\tau_a$):
\begin{equation}
	\dot n_{f} = \frac{1-n_{f}}{\tau_d} - \frac{n_{f}^2}{\tau_a}.
	\label{eqn:ChemEqSupM}
\end{equation}
At thermal equilibrium (in the absence of external field), the ratio of these two \rev{times} is given by $e^{2 \mu_e}$, as discussed in the previous section. {When driven} out of equilibrium however, these two quantities may depend on the applied electric field $\vecb E = E \, \unitary x$ {acting along $x$}.

Onsager showed that it is possible to link both $\tau_a$ and $\tau_d$ to the anion-cation correlation function $g$. Assuming that a cation is held fixed at the origin, $g(r,\theta)$ is the probability density of finding a negative ion at polar position $(r,\theta)$. It follows a Smoluchowski equation:
\begin{equation}
	\partial_t g = 2D\vecb {\nabla} \cdot ( \vecb {\nabla} g + g \vecb \nabla V ),
	\label{eq:Smoluchowski}
\end{equation}
where $V$ is the total (dimensionless) electrostatic potential at $(r,\theta)$. It reads:
\begin{equation}
	V =  \frac{1}{T^*}\log\frac{r}{r_0} - \frac{r \cos \theta}{\lE}.
	\label{eqn:PhiOnsager}
\end{equation}
\revv{Here, we made the approximation that $\log r/(r + \xi) \sim \log r$, which is reasonable for $r < \xi$. Since $\xi = 14\,$nm, this approximation is valid over the typical lengthscales of ionic correlations.} The first term of the potential corresponds to the unscreened interaction of two ions in confinement. This assumption is valid if the system is sufficiently paired up ($T^* < T^*_c$) so that the influence of other free ions can be neglected. Quantitatively, the Debye length $\lD$ diverges if the system is fully paired, and therefore is not a relevant length scale. The second term corresponds to the external field, characterized by the following length scale:
\begin{equation}
	\lE = \frac{k_B T}{Ze {|E|}}.
\end{equation}
The potential $V$ has a maximum for $r \sim \lE/T^*$; a pair can be expected to break if its two ions are separated by a larger distance. This fixes the typical length scale of correlations in presence of the field. 

Assuming the system has reached a steady state, we obtain:
\begin{equation}
	\left[\Delta + \left(\frac{1}{T^*r} - \frac{\cos \theta}{\lE}\right)\partial_r + \frac{\sin \theta}{r \lE} \partial_\theta \right] g = 0.
\end{equation}
We then perform the change of variable $\vecb r \to \vecb u = \vecb r/\lE$:
\begin{equation}
	\left[\Delta + \left(\frac{1}{T^*u} - \cos \theta\right)\partial_u + \frac{\sin \theta}{u} \partial_\theta \right]g = 0.
	\label{eqn:SmolReduced}
\end{equation}
This shows that the problem is scale invariant, as it is now entirely determined by a single dimensionless parameter $T^*$; this property is unique to the 2D geometry (where the Bjerrum length is infinite). Onsager's trick consist in splitting the correlation function into two parts:
\begin{equation}
	g = g_d + g_a,
\end{equation}
where $g_d$ and $g_a$ are two solutions of \eqref{eqn:SmolReduced} associated with a source or a sink of particles at the origin, respectively:
\begin{align}
	\int_{0}^{2 \pi }-2D\left[\vecb {\nabla} g_a + g_a \vecb \nabla V \right] \cdot \unitary r \, r \di \theta = -C,\\
	\lim\limits_{r \to \infty}g_a =\rho,
\end{align}
and
\begin{align}
	\int_{0}^{2 \pi }-2D\left[\vecb {\nabla} g_d + g_d \vecb \nabla V \right] \cdot \unitary r \, r \di \theta = +C,\\
	\lim\limits_{r \to \infty}g_d =0,
\end{align}
where $C$ is a positive constant independent of $r$ \rev{and $\rho$ is the average ionic density far from the central ion}. In other words, $g_a$ describes a background of free ions recombining with the central ion to form new pairs, and $g_d$ pairs that break under the electric field. These two functions will allow us to compute $\tau_a$ and $\tau_d$.

Since any constant is a solution of the Smoluchowski equation, it is easy to see from the boundary condition that:
\begin{equation}
	g_a = \rho,
\end{equation}
and straightforward integration yields:
\begin{equation}
	C = \frac{4 \pi D \rho}{T^*}.
\end{equation}
This is a recombination rate, defining the pair association time:
\begin{equation}
	\tau_a = \frac{T^*}{4 \pi D \rho}.
	\label{eqn:TauA}
\end{equation}
Interestingly, we find that the formation time of pairs is independent of the electric field. This result is general and is also valid in the bulk (but not in 1D), as recombination of freely diffusive ions is an uncorrelated process.

Onsager's original approach to compute $g_d$, which involves an arduous expansion in terms of special functions, strongly relied on a fact that the 3D Smoluchowski equation is spatially separable; it fails in our 2D case. Instead, we exploit the fact that the 2D Smoluchowski equation is scale-invariant, and show how all relevant quantities can be computed up to a geometrical factor.

\subsubsection{Self-similarity of the correlation function}
We start be noticing that another solution of the Smoluchowski equation \eqref{eqn:SmolReduced} is the Boltzmann distribution:
\begin{equation}
	g_0(u, \theta) = \exp\left[- \frac 1 {T^*} \ln \frac{u\lE}{r_0} + u \cos \theta \right].
\end{equation}
This is not the correct solution of the problem however, as the Boltzmann distribution only makes sense at thermal equilibrium. In particular, it does not verify the correct boundary conditions at $r\to \infty$. However, for $u \ll 1$ the external field is negligible compared to the field created by the central ion and pairs are in quasi-equilibrium (ions are strongly correlated and remain bounded over long timescales). \rev{Therefore, the total correlation function $g$ should diverge like $g_0$ for $u \to 0$. Since $g_a$ is bounded around $u = 0$, this condition also provides us a boundary condition for $g_d$, which is therefore the solution of:} 
\begin{equation}
	\left[\Delta + \left(\frac{1}{T^* u} - \cos \theta\right)\partial_u + \frac{\sin \theta}{u} \partial_\theta\right] g_d= 0,
\end{equation}
\begin{equation}
	g_d \sim K_a \left(\frac{\lE}{r_0}\right)^{-1/T^*}u^{-1/T^*} \ \text{for} \  u \to 0,
\end{equation}
\begin{equation}
	\lim\limits_{u \to \infty } g_d(u) = 0,
\end{equation}
where $K_a$ is a constant determined by the fact that the flux of $g_d$ should be equal to $+ C$. The solution to this problem is unique because $g_d$ is known on the whole boundary of the domain. $\lE$ now only appears in a boundary condition at $u=0$, and the system is linear, so $g_d$ is fully determined from a single scaling function:
\begin{equation}
	g_d(u) =  K_a \left(\frac{l_E}{r_0}\right)^{-1/T^*}G(u),
\end{equation}
where $G$ is a function that depends only on $T^*$. The balance between the fluxes of $g_d$ and $g_a$ reads:
\begin{equation}
	2DK_a \left(\frac{\lE}{r_0}\right)^{-1/T^*} \mathcal{F}= \frac{4 \pi D \rho}{T^*},
\end{equation}
where $\mathcal F$ is the flux of the function $G$:
\begin{equation}
	\mathcal F = -\int_{0}^{2 \pi }\left[\vecb {\nabla} G + G \vecb \nabla \hat V \right] \cdot \unitary u \, u \di \theta,
\end{equation}
where the dimensionless potential $\hat V$ is given by:
\begin{equation}
	\hat V = \frac{1}{T^*}\log u - u \cos \theta.
\end{equation}
The flux $\mathcal F$ has the dimension of an inverse length squared and is independent of $\lE$ or $\rho$, since it describes a single pair breaking event. As the only remaining length-scale in the problem is $r_0$, we have (up to a geometrical factor):
\begin{equation}
	\mathcal F \simeq r_0^{-2}.
\end{equation}
We obtain the association constant $K_a$:
\begin{equation}
	K_a = 2 \pi \left(\frac{\lE}{r_0}\right)^{1/T^*} \frac{\rho r_0^2}{T^*}=\frac{\tau_d}{\tau_a},
\end{equation}
and the dissociation time $\tau_d$:
\begin{equation}
	\tau_d = \frac{r_0^2}{2 D}\left(\frac{\lE}{r_0}\right)^{1/T^*}.
	\label{eqn:WienTauD}
\end{equation}
In the steady state, the free ion fraction $n_{f}$ is given by Equation \eqref{eqn:ChemEqSupM}:
\begin{equation}
	n_{f} = \frac{\tau_a}{2\tau_d}\left(\sqrt{1 + \frac{4 \tau_d}{\tau_a}} - 1\right).
	\label{eqn:nfIsolatedPair}
\end{equation}
\rev{It should be noted that this result is general for any system with Onsager's pairing kinetics, and does not depend on details of ionic interactions, as long as $\tau_a$ and $\tau_d$ are known. In the limit of weak applied field, $\tau_d \gg \tau_a$ and $n_f \ll 1$. We obtain:
\begin{equation}
	n_f \simeq \sqrt{\frac{\tau_a}{\tau_d}},
\end{equation}
with $\tau_a$ and $\tau_d$ given by Eqs. \eqref{eqn:TauA} and \eqref{eqn:WienTauD} respectively. Assuming each free ion contributes linearly to conduction, $j(E) \propto E \times n_f(E)$} and we obtain the ionic current due to the Wien effect:
\begin{equation}
	|j(E)| \propto {|E|^{1+1/2T^*}}.
\end{equation}
{This predicts there that the ionic  conductivity scales sublinearly with the temperature, as {$\sigma \sim  |E| ^{1/2T^*}$.}}
While this prediction reproduces qualitatively the non-linearity observed in {the simulations}, it typically underestimates the conductivity of the system by more than one order of magnitude. In particular, we do not find the exponent {measured in the simulations}.

In the next section, {in order to explain this this behavior, we now take into} account the formation of larger ionic clusters in presence of an external field. We do so by first rederiving our result through scaling arguments, and then amending it using the same scaling logic.

\rev{Lastly, the above derivation critically relies on the (quasi-)logarithmic shape of the interaction potential: because its typical lengthscale $\xi$ is much larger than any other physical lengthscale in the problem, ionic correlations become self-similar. This situation is unique to the 2D case; for example, in bulk electrolytes, ions interact through the usual $1/r$ Coulomb's law (which has a typical lengthscale set by the Bjerrum length $\lB$, see Section II.A.2) and computing correlations is much more mathematically involved \cite{onsager_deviations_1934}.}

\subsubsection{Scaling laws and anisotropic conduction}

In the above derivation, the key assumption was the divergence of the Debye length $\lD$, so that the problem becomes self-similar. We recall that $\lD \propto n_f^{-1/2}$, and we found that $n_f \propto E^{1/2T^*}$, so that $\lD \propto E^{-1/4T^*}$. If $T^* < 1/4$, then $\lD \gg \lE$ for all relevant values of the electric field, since $\lE \propto E^{-1}$.

This scaling argument seemingly validates our approach of considering that $\lE$ sets the scale of all correlations between ions. However, this argument only makes the sense when considering correlations along the $x$ axis: ions separated by more than $\lE$ are carried away by the electric field, so correlations on larger scale may be neglected. This is not the case, however, along the $y$ axis, so actually one should not neglect the influence of $\lD$ on the problem: correlations are anisotropic.

The above analysis can be supported qualitatively by analyzing simulation results. In Brownian dynamics, we observe that ions tend to form elongated clusters in the direction of the electric field. Ions are able to move within these clusters, but remain within them for long times (see Figure S3). This suggests that the mobility of ions in the direction $x$ of the electric field may increase with the field strength, while diffusion in the orthogonal direction $y$ remains constrained by electrostatic correlations. 

Taking anisotropic correlations into account, however, requires to account for many-body interactions in the Smoluchowski equation for $g$, which makes the problem intractable. Instead, we use a scaling argument, that we now detail.

Let us start by recovering the previous result using scaling laws. Equation \eqref{eqn:WienTauD} can be recast as an Arrhenius law:
\begin{equation}
	\tau_d = \tau_{\text{diffusion}}\exp\left[- 2 \beta \Delta F\right] = \frac{r_0^2}{2D}\left(\frac{\lE}{r_0}\right)^{1/T^*},
\end{equation}
where $\Delta F$ is the ``free energy barrier'' to break a pair. It reads:
\begin{equation}
	\beta \Delta F \simeq - \frac{\log \lE /r_0}{2T^*}.
\end{equation}
This argument is not fully rigorous, as the problem is far from equilibrium and this energy barrier strongly depends on an out-of-equilibrium quantity (the length scale $\lE$). However, this last expression is identical, upon replacing $\lE$ by $\lD$ to the (properly defined) free energy cost of breaking a pair at equilibrium, given by Equation \eqref{eqn:FreeEnergyCost}. We thus find that the \textit{kinetic} energy barrier to break a pair is similar to the \textit{thermodynamic} energy gap between the paired and the unpaired states, if we admit that the typical size of the ionic atmosphere in presence of an external field is given by $\lE$ instead of the Debye length $\lD$. 

Let us now use this simple argument to account for anisotropic correlations. Since the typical size of the correlation cloud of is $\lE$ along the $x$ axis and $\lD$ along the $y$ axis, {one may roughly approximate} its overall spatial extension as $\ell = \sqrt{\lD \lE}$. In this case, the free energy barrier to breaking a pair should read:
\begin{equation}
	 \beta \Delta F \simeq - \frac{\log \lE/r_0}{4T^*} - \frac{\log \lD/r_0}{4T^*},
\end{equation}
so that it is now the sum of two terms, corresponding to the energy barrier that an ion has to overcome to escape a cluster in the $x$ or $y$ direction, respectively. Conduction in the direction of the field is therefore associated with the Arrhenius timescale corresponding to the first term in the free energy:
\begin{equation}
	\tau_{d,x} = \frac{r_0^2}{2D}\left(\frac{\lE}{r_0}\right)^{1/2T^*},
\end{equation}
and we can define the proportion $n_x$ of ions that can freely move along the $x$ axis. It follows a evolution equation similar to Equation \eqref{eqn:ChemEqSupM}, with $\tau_{d,x}$ replacing $\tau_d$. We obtain:
\begin{equation}
	n_x = \frac{\tau_{a}}{2\tau_{d,x}}\left(\sqrt{1 + \left(\frac{2 \tau_{d,x}}{\tau_{a}}\right)^2} - 1\right) \underset{E\to 0}{\propto} E^{1/4T^*}.
	\label{eqn:ClusterizedNf}
\end{equation}
We finally obtain the following prediction for the total ionic current:
\begin{equation}\label{eq:jofEth}
	|j(E)| \propto |E|\times n_x(E) \propto {\times|E|^{\alpha(T^*)}},
\end{equation}
with
\begin{equation}
	\alpha(T^*) = 1 + \frac{1}{4 T^*}.
	\label{eqn:alpha}
\end{equation}

\subsubsection{Comparison with simulations and discussion}

\begin{figure}
    \centering
    \includegraphics{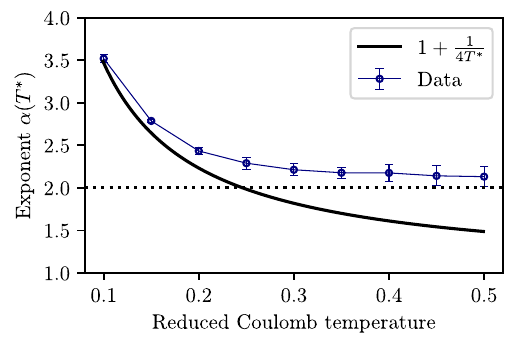}
    \caption{\label{fig:exponent}Symbols: Exponent $\alpha$ defined by the non-linearity in the IV curve, $|j(E)| \propto |E|^\alpha$. Black solid line: Theoretical prediction of the 2D Wien effect, see Equation \eqref{eqn:alpha}. Dotted line: bulk exponent $\alpha = 2$, obtained through Onsager's theory of the Wien effect \cite{onsager_deviations_1934}.}
\end{figure}

{The comparison between our BD simulations and Equation \eqref{eq:jofEth} is plotted in Figure \ref{fig:exponent}.a and b.} Below the Kosterlitz-Thouless transition, the agreement is quantitative and the observed exponent $\alpha(T^*)$ in simulations matches the theoretical prediction. \rev{Conductivity scales like:
\begin{equation}
	{\text{For } T^* < T_c, \quad  \sigma(E) \propto |E|^{\alpha - 1}, \quad \alpha \simeq 1 + \frac{1}{4 T^*}.}
\end{equation}}
Interestingly, this exponent is non-universal as it strongly depends on temperature; this contrasts with bulk electrolytes where the second Wien effect results in a universal conducivity increment scaling like $\delta \sigma \propto |E|$

Above the pairing transition, however, we observe deviations to the law. We find that the conductivity increases linearly with the applied field (Fig. 3b), which echoes the bulk Wien effect as mentioned above. We now suggest a possible explanation. 

The key element in the above derivation that lead to the non-universal exponent $1/4T^*$ was the self-similarity of the correlation function. Below the KT transition, this assumption is valid as almost all ions are paired up, resulting in a diverging correlation function for electrostatic correlations: $\lD \to \infty$. However, for $T^* > T_c$, some free ions remain even for $E = 0$, and $\lD$ always remains finite. Therefore, the correlation function is not self-similar as the problem now possesses two typical lengthscales: $\lD$ and $\lE$. The second Wien effect may then be obtained from scaling laws: in absence of field, an ion pair breaks when its two ions are separated by more than $\lD$ (after which they cease to interact). Under an electric field, this transition state between pairs and free ions is destabilized by roughly a factor $\lD/\lE \propto |E|$. Since the dissociation timescale is approximatively the Arrhenius time associated with this intermediate state, conductivity also increases by a factor linear in $|E|$: we recover the scaling of Onsager's Wien effect in bulk electrolytes. We therefore obtain:\rev{
\begin{equation}
	{\text{For } T^* > T_c, \quad  \sigma(E) - \sigma(0) \propto |E|^{\alpha-1}, \quad \alpha \simeq 2,}
\end{equation}}
in good agreement with numerical simulations (Fig. 5b).

It should be noted, however, that in the bulk, the remaining free ions at $E = 0$ do not play a significant role at sufficient dilution, because $\lD > \lB$ and so correlations still have a typical length scale $\lB$. In other words, the conductivity increment will scale like $E$ at all temperatures and does not directly depends on ion concentration, unlike in confined electrolytes.

The above derivation was performed assuming that all ions are paired up. In what follows, we account for ions that remain free even in absence of any field by adding a contribution $n_f(0)$ to the above result; this term is determined by the procedure described in SI.
The above sections were dedicated to the theoretical analysis of the 2D Wien effect at low temperature ($T^* < T^*_c$). We find that the conductivity of 2D confined electrolytes evolves as a power law of the electric field, with a non-universal exponent $1/4T^*$. This contrasts with the {3D} bulk case, where the {Wien} exponent is 1. 

Lastly, we note that in both cases $T^* > T_c$ and $T^* < T_c$ the exponent cannot be found through simple symmetry arguments, which would dictate that $\sigma \propto E^2$ (as the system is invariant by reversing the direction of the electric field). This originates in the fact that the correlation function $g$ becomes strongly polarized in the direction of the field, breaking the $x \to -x$ symmetry. In the bulk case, the conductivity increment scales like the ratio $\lB/\lE \propto E$, with $\lB$ the Bjerrum length. Since this quantity is infinite in 2D, the problem becomes self-similar and the increment is found to scale like a non-universal power law.

\section*{Conclusion}
In this paper, we investigate the effect of long range electrostatic correlations on the equilibrium and transport of ions confined in a 2D slit. We use a combination of molecular dynamics simulations, analytical theory and Brownian dynamics (where water and channel walls are treated implicitly, and ion-ion interactions are renormalized). In all cases, we find that 2D confinement results in stronger electrostatic interactions, leading to the formation of ionic pairs. We showed that this phenomenon is associated to a phase transition analogous to the Kosterlitz-Thouless transition, and suppresses linear ionic conduction at low temperature.

In addition, the application of an external field can result in the breaking of ion pairs and in an increase in conduction. This process, known as the Wien effect, leads to strongly non-linear ion transport under confinement.

We expect that the effective potential approach used in this paper could be extended to explore other materials\cite{kavokine_interaction_2022}, but also other geometries, for example the case of multiple ionic layers\cite{coquinot_collective_2024}, by changing the effective interaction potential accordingly. \rev{Models of ionic clusters dynamics are also useful to study ionic liquids, where many parallels can be drawn for the relations between ion correlations and transport phenomena \cite{feng2019free}.}

Overall, we obtain excellent agreement between our analytical models and numerical results, for both the pairing transition and the 2nd Wien effect. In particular, we find that in the pair-dominated regime, the ionic current behaves like a power law of the applied field, with a non-universal exponent that can be predicted from analytical field theories. {This regime, where the conductivity strongly vanishes at low electric field, can be considered as ionic coulomb blockade situation -- although no gating dependence is considered here.}

This work is {also a further demonstration} of the very particular nature of electrostatic interactions in confined geometry. The strong ionic correlations gives rise to a complex variety of structures and behaviours, but the large-scale picture remains unaffected and can be adequately understood in term of a small set of parameters. Consequently, this work sheds light on the structure of ion-ion correlations in confined systems, and on the ionic dynamics of nanofluific in general.

{Finally, the emergence of strongly non-linear conduction effects in 2D is a richness which can be exploited to develop nanofluidic systems with advanced properties, such as memristors \cite{robin2023long,robin_nanofluidics_2023}.
This is an opportunity which will certainly result in further developments in this active domain \cite{kamsma2024brain,emmerich2024nanofluidic,paulo2023hydrophobically}} 

\rev{
\section{Supplementary Material}
Supplementary Material includes details of numerical simulations, including the values of all numerical parameters. We also detail there the procedures to accurately extract the average number of pairs from numerical correlation functions, as well as the IV curve exponent $\alpha$.
}

\section*{Acknowledgements}
The authors thank B. Coquinot and G. Monet for fruitful discussions. LB acknowledges support from ERC-Synergy grant agreement No.101071937, n-AQUA. PR acknowledges support from the European Union’s Horizon 2020 research and innovation program under the Marie Sklodowska-Curie grant agreement No.101034413.

\bibliography{ms.bib}
\end{document}


\title{Supplementary Informations: Ionic association and Wien effect in 2D confined electrolytes}

\author{Damien Toquer}
\affiliation{Laboratoire de Physique de l'Ecole Normale Sup\'erieure, 24 rue Lhomond, 75005, Paris, France}
\author{Lyd\'eric Bocquet}
\email{lyderic.bocquet@ens.fr}
\affiliation{Laboratoire de Physique de l'Ecole Normale Sup\'erieure, 24 rue Lhomond, 75005, Paris, France}
\author{Paul Robin}
\email{paul.robin@ist.ac.at}
\affiliation{Institute of Science and Technology Austria, Am Campus 1., 3400 Klosterneuburg, Austria}

\pacs{}

\maketitle

\pagebreak
\onecolumngrid

\tableofcontents

\setcounter{equation}{0}
\setcounter{figure}{0}
\setcounter{table}{0}
\setcounter{page}{1}
\setcounter{section}{0}
\makeatletter
\renewcommand{\theequation}{S\arabic{equation}}
\renewcommand{\thefigure}{S\arabic{figure}}
\renewcommand{\thetable}{S\arabic{table}}
\renewcommand{\thesection}{S\arabic{section}}

\section{Simulation details}\label{sec:simu_detail}
Simulations are carried out using GROMACS\cite{abraham_gromacs_2024} and LAMMPS\cite{thompson_lammps_2022} softwares, {respectively}. Both equilibrium (without any external electric field) and non equilibrium simulations (with an external electric field) are used. The trajectory are obtained from simulation results using the MDAnalysis\cite{gowers_mdanalysis_2016} Python library, and visualized with Ovito\cite{stukowski_visualization_2010}. {Structure analysis of ion clusters} is done using the network analysis Python library NetworkX \cite{osti_960616}.

\subsection{Detail for BD simulations}
In Brownian dynamics (BD) simulations, ions evolve in a 2D continuous space representing the nanochannel. The following equation of motion is solved for each ion $i$:
\begin{equation}
    \frac{\text{d} \vec{r}_i}{\text{d} t} = \frac{D}{k_BT}\left(z_ieE\vec{u}_x-\vec{\nabla} U_i\right)+\sqrt{2D}\vec{\eta}_i(t),
\end{equation}
where $z_i = Z$ for cation and $z_i = -Z$ for anions, $E$ is the external field, $D$ is the diffusion coefficient (identical values for both cations and anions), $U_i$ is the total potential felt by the ion i, {and the $\vec{\eta}_i$ are independent 2D white noise forces:}
\begin{equation}
    \langle \vec{\eta}_i(t) \rangle = \vec{0}, \quad\langle \vec{\eta}_i(t)\cdot \vec{\eta}_j(t') \rangle = \delta_{ij}\delta(t-t')
\end{equation}
\rev{We assume that both types of ions have the same diffusion coefficient. To account for situations where this might not be the case, $D$ can be taken as the average of the coefficients of anions and cations.}

\rev{In our BD simulations, the temperature is used as a control parameter tuning the relative strength of ionic interactions. However, this also impacts the relaxation time of Brownian dynamics $\tau = Dm/k_BT$. We choose to simplify the analysis by assuming that this relaxation time is constant, which amount to adjust the diffusion coefficient with temperature such that:
\begin{equation}
\tau =\frac{Dm}{k_BT} = 9.6\ \text{fs}
\end{equation} 
We choose $m = 23\ \text{uma}$ for both ions.}

The interaction is given by the following interaction potential derived in the main text Section I:
\begin{equation}\label{eq:interpot}
    \beta V_{ij}(r_{ij}) = -\frac{\text{sign}(z_iz_j)}{T^*}\log\left(\frac{r_{ij}+C}{r_{ij}+\xi}\right),
\end{equation}
where $z_i=\pm 1$ is the charge of the ion $i$, $r_{ij}$ is the distance between the ions and $\beta = 1/k_BT$. To avoid a divergence of the potential at short ionic distances - without requiring additional repulsive potential - the effective potential derived in the last section is shifted by a short range cutoff $C = 3\AAA$. This enables the use of a large time step of 5 ps. The interaction between the ions can be controlled by changing either the charges of the ions, or the temperature. In relevant experimental conditions, $T$ is usually fixed to 300K, and $Z$ is varied by using salts of various valence, and $T^* = 0.1$ corresponds to divalent ions and $T^* = 0.4$ to monovalent ions. For practical reasons, and to be able to continuously vary the interaction strength, we rather fix in this work $Z=2$ and vary the temperature $T$.

The total effective interaction potential felt by an ion in Brownian dynamics simulation is finally given by:
\begin{equation}
    U_i = \sum_{j\neq i}V_{ij}^c(r_{ij}),
\end{equation}
where $V_{ij}^c$ is the cut off version of the effective potential. It is defined by:
\begin{equation}
    V_{ij}^c(r_{ij}) = \left\{
        \begin{array}{cc}
            V_{ij}(r_{ij}) & \quad\text{if }r_{ij}<r_c \\
            0 & \quad\text{otherwise}
        \end{array}
    \right.,
\end{equation}
with $r_c = 20$ nm.

The system consists of $N_+ = N_- = N$ ions in a 2D slit of dimension $L \times w$. In section II, we use $N=100$ ions, in a box of lendth and width $L = w = 100$\ nm, for a total simulations times of $10^6$ timesteps. In section III, we use $N=4000$ ions, in a box of length $L = 2000$ \ nm and width $w=200$ \ nm, for a total simulations times of $10^5$ timesteps. For all BD simulations, we have an ionic concentration of $10^{16} \ \text{atom}.\text{m}^{-2}$, or 0.025 M.

\subsection{Detail for MD simulations}
Molecular dynamics simulations consist of two sheets of fixed carbon atoms, with periodic boundary condition. In order to reproduce the crystalline structure of graphene, a rectangular unit cell of size $\sqrt{3}d_c\times 3d_c$ is created, corresponding to 4 carbon atoms, where $d_c=1.42\ \AAA$ is the carbon-carbon distance (see Figure 1 of main text for the definition of the unit cell). This cell is replicated 81 times along the $x$ direction and 46 times along the $y$ directions for each sheet. Both sheets are separated by a distance $h$, which creates a slit of height $h=7\ \AAA$. Because of periodicity, we extend the simulation box to 20nm in the z direction to avoid interaction between periodically replicated images. This gives a total simulation box of size $19.9\ \text{nm}\times 19.6\ \text{nm}\times 20\ \text{nm}$. The inside of the slit is {filled with water ad ions} consisting in 2 identical mono-atomic species with a varying charge $\pm Ze$. 
The number of water molecules and ions are respectively $N_w = 4173$ and $N_+=N_-=N=100$. This corresponds to a ionic concentration inside the slit of 0.6 M. The procedure used to estimate the number of water molecules in this geometry is presented in Section \ref{sec:waterest}.

Water molecules are modeled using the SPC/E model\cite{berendsen_missing_1987} (3 point charges model) and maintained rigid with the SHAKE algorithm\cite{ryckaert_numerical_1977}. Short-range interactions are modeled using Lennard-Jones (LJ) potentials. The cut-off distance for LJ and Coulombic interactions is 1.2 nm. The atom mass, charges and Lennard Jones parameters are summarized in Table S1. For both ions we use the same LJ parameters and masses (which are taken from the sodium force-field) with opposite charges. Long-range Coulombic interactions are treated with the smooth particle-mesh Ewald method\cite{darden_particle_1993,essmann_smooth_1995}. When initializing the simulation, the total energy is minimized using the steepest descent minimization algorithm.

Simulations are performed in the NVT ensemble using the velocity Verlet\cite{swope_computer_1982} algorithm with a timestep $\text{d}t = 1$ fs. The system is coupled to a thermostat at 300K using velocity rescaling with a stochastic term\cite{bussi_canonical_2007}. The simulation runs for $1.5\times 10^7\ \text{d}t = $15 ns, and the first 5 ns are discarded for thermalization during analysis.

\begin{table}[H]
\begin{center}
\begin{tabular}{c||c|c||c|c||c}
Atom type & Mass (amu) & Charge (e) & $\varepsilon$ (kcal/mol) & $\sigma$ (\AA) & Source\\
\hline
\hline
O & 15.999 & -0.8476& 0.1553 & 3.166 & SPCE\cite{berendsen_missing_1987}\\
H & 1.008 & 0.4238 & 0 & 0 & SPCE\cite{berendsen_missing_1987}\\
\hline
C & - & 0 & 0.0567 & 3.214 & Werder\cite{werder_water-carbon_2008}\\
\hline
Na & 22.990 & $\pm z_i$ & 0.1076 & 2.310 & Loche\cite{loche_transferable_2021}\\
\end{tabular}
\caption{MD simulation parameters used in this study. As this work mainly focuses on electrostatic interactions, we use a fictional perfectly symmetric salt where the anion has the same interaction paremeters as the cation, albeit with an opposite charge. We therefore use the parameters of sodium ions in both cases.}
\end{center}
\label{tab:LJparams}
\end{table}

\subsection{Determination of water density in confinement}\label{sec:waterest}
The following procedure is carried out using the LAMMPS\cite{thompson_lammps_2022} software in order to estimate this quantity. A slit of height $7\ \AAA$ is connected to 2 reservoirs, each pressurised by a piston, at a pressure $P=1$\ atm. Both pistons are made of fictive atom of mass 18 uma, no charge, and the LJ parameters of oxygen, placed on a centered rectangular lattice. We then compute the mean number of water molecules in the slit during a 20 ps simulations. We obtain a final density of $0.112$\ \AA$^{-2}$. {This result is in agreement with previous works \cite{yoshida_dripplons_2018}}. In the presence of ions we replace $2N$ water molecules by $N$ cations and $N$ anions.

\section{Determination of free charge carrier fraction at equilibrium in simulations}

\subsection{Convention for correlation functions}

The total pair correlation function is defined as:
\begin{equation}
    g_p(\vec{r}) = \frac{\mathcal{A}}{4N^2}\left\langle\sum_{i}^{2N}\sum_{j\neq i}^{2N}\delta(\vec{r}-(\vec{r}_i-\vec{r}_j))\right\rangle,
\end{equation}
where $\mathcal{A}=Lw$ is the area of the slit, $\vec{r}_i$ is the position of ion $i$, and where $\langle\cdot\rangle$ denotes an ensemble average. Without external field, the system is symmetric by rotation, and the radial pair correlation functions can be used instead:
\begin{equation}
    g_p(r) = \frac{1}{2\pi r} \frac{\mathcal{A}}{4N^2}\left\langle\sum_{i}^{2N}\sum_{j\neq i}^{2N}\delta(r-|\vec{r}_i-\vec{r}_j|)\right\rangle.
\end{equation}
We can alternatively define $g_{\text{op}}$ and $g_{\text{id}}$ the pair correlation functions for respectively oppositely charged and identically charged ions :
\begin{equation}
    \begin{array}{ccc}
         g_{\text{op}}(r) &=& \frac{1}{2\pi r} \frac{\mathcal{A}}{N^2}\left\langle\sum_{i,\text{cations}}^{N}\sum_{j,\text{anions}}^{N}\delta(r-|\vec{r}_i-\vec{r}_j|)\right\rangle \\
         g_{\text{id}}(r) &=& \frac{1}{2\pi r} \frac{\mathcal{A}}{N^2}\left\langle\sum_{i,\text{cations}}^{N}\sum_{j\neq i,\text{cations}}^{N}\delta(r-|\vec{r}_i-\vec{r}_j|)\right\rangle
    \end{array}
\end{equation}
$g_{\text{op}}$ is invariant under inversions of anions and cations, however $g_{\text{id}}$ is slightly asymmetric - even for symmetric anion and cation - because of the structure of water molecules. In the following we take the average of both anion-anion and cation-cation pair correlation functions.

In practice, we compute all the $2N^2N_t$ unique ionic distances, where $N_t$ is the number of simulation steps used, and $N$ is the number of ions. The distance between ions are corrected to take into account the periodicity. 
{Radial pair histograms} are computed every 5 ns ($10^{3}$ timesteps), for a total simulations length of 5 $\mu$s ($10^6$ timesteps), and the first 500 ns are discarded. In order to estimate the uncertainty on those quantity, we split the time interval into 10 windows, and we compute the standard deviation of the window averages.

The running coordination number $\nop(r)$ and $\nid(r)$ can be obtained upon integration of the radial pair correlation function of respectively opposite and identical ions. It corresponds to the number of ions in a shell of radius $r$ around any ions:
\begin{equation}\label{eq:coordinationdef}
    \nop(r) = 2\pi\frac{N}{\mathcal{A}}\int_0^r \gop(u)u\text{d}u.
\end{equation}
\begin{equation}
    \nid(r) = 2\pi\frac{N}{\mathcal{A}}\int_0^r \gid(u)u\text{d}u.
\end{equation}
The counter charge density around an ion can be defined as:
\begin{equation}
    q(r) = -\frac{1}{4\pi rN}\left\langle\sum_{i}^{2N}\sum_{j\neq i}^{2N}\text{sign}(z_iz_j)\delta(r-|\vec{r}_i-\vec{r}_j|)\right\rangle,
\end{equation}
$q$ can be integrated to obtain:
\begin{equation}
    Q(r) = 2\pi\int_0^r q(u)u\text{d}u = \nop(r)-\nid(r),
\end{equation}
which corresponds to the total counter charge in a disk of radius $r$ around an ion. Those functions are normalized by the central ion charge, such that $Q(0) = 0$, and $Q(r\to\infty) = 1$, which corresponds to global neutrality of the system.

Adding an external electric field breaks the isotropy of the system. In this case, the running coordination number $N(r, E)$ can still be defined, but it does not exactly have the same physical meaning, as the anisotropic correlation are integrated on a circle.

\subsection{Free ions fraction in BD simulations}\label{sec:nf_bd}
\subsubsection{Pairing criterium}

\rev{In the next paragraph, we develop a method to estimate the value of the free ion fraction in the absence of external field. We will make use of an alternate definition of ion pairs than Equation (9) of main text, based on correlation functions. We show that this definition is robust regardless of the magnitude of the external field. We will therefore use it as a proxy to estimate the free ion fraction at equilibrium.}

We define a typical length scale $d(E)$ associated with the formation of pairs through the correlation function. We assume a separation of the ions into 2 different types of cluster, the free ions and pairs. By splitting the sum in the definition of $\gop(r)$ between paired and unpaired ions, \rev{we can define:
\begin{equation}
    g_{\text{intra}}(r) = \frac{1}{2\pi r}\frac{\mathcal{A}}{N^2}\left\langle\sum_{\substack{i,\text{anions} \\ \text{paired}}}^N\sum_{\substack{j,\text{cations} \\ \text{paired with i}}}^N\delta(r-r_{ij})\right\rangle,
\end{equation}
and:
\begin{equation}
    g_{\text{inter}}(r) = \frac{1}{2\pi r}\frac{\mathcal{A}}{N^2}\left\langle\left(\sum_{\substack{i,\text{anions} \\ \text{paired}}}^N\sum_{\substack{j,\text{cations} \\ \text{other}}}^N+\sum_{\substack{i,\text{anions} \\ \text{unpaired}}}^N\sum_{j,\text{cations}}^N\right)\delta(r-r_{ij})\right\rangle.
\end{equation}}

\rev{We observe that:
\begin{equation}
    2\pi\frac{N}{\mathcal{A}}\lim_{r\to\infty}\int_0^rg_{\text{intra}}(u)u \text{d}u = n_p,
\end{equation}
and:
\begin{equation}
    2\pi\frac{N}{\mathcal{A}}\lim_{r\to\infty}\int_0^rg_{\text{inter}}(u)u\text{d}u = N-n_p,
\end{equation}
where the half factor is the consequence of the fact that only half of the free ions are anions. We finally define the following functions:
\begin{equation}
    \rho_{\text{intra}}(r) = \frac{1}{n_p}\times 2\pi r\frac{N}{\mathcal{A}}g_{\text{intra}}(r),
\end{equation}
and:
\begin{equation}
    \rho_{\text{inter}}(r) = \frac{1}{N-n_p}\times 2\pi r\frac{N}{\mathcal{A}}g_{\text{inter}}(r).
\end{equation}
Then by definition:
\begin{equation}
    N_{\text{op}}(r) = n_p\int_0^r\rho_{\text{intra}}(u)\text{d}u+(N-n_p)\int_0^r\rho_{\text{intra}}(u)\text{d}u,
\end{equation}
with:
\begin{equation}
    \int_0^{\infty}\rho_{\text{intra}}(u)\text{d}u = \int_0^{\infty}\rho_{\text{inter}}(u)\text{d}u = 1.
\end{equation}}

$\rho_{\text{intra}}$ and $\rho_{\text{inter}}$ are respectively the correlation functions of ions within a same cluster (short distances) and correlations of ions in different clusters (large distances). Both of them vary with the external field. When the overlap between those distributions is small, we can define an intermediate length $d(E)$ between intra and inter cluster correlation, which corresponds to the maximal size of a pair. When $r \simeq d(E)$, the first integral is close to 1, when the second one remains close to 0. This yields:
\begin{equation}
    \nop(d(E), E) = n_p(E).
\end{equation}
In other words, $\nop$ exhibits a plateau (see Figure S2), associated to a minimum of $\gop$. For $T^*\gtrsim 0.1$, we do not observe any minima in the correlation functions, which indicates that a geometric criterion is not adapted for our system. Yet, as the fraction of charge carrier is still well defined in presence of external field, we can use this number to define a typical distance between paired ions $d(E)$:
\begin{equation}
    \nop(d(E>0), E>0) = 1-n_f(E>0),
\end{equation}
where $n_f(E>0)$ is computed from transport simulations using Equation (9) of main text. 

\subsubsection{Free ion fraction}

Once $d(E)$ is computed, we extrapolate the value of the pair distance without external field $d(E=0)$, by assuming that the pair distance decays exponentially with the field. This function is used instead of a linear one as it fits better the whole field dependency of the pair distance. We can finally estimate the free ion fraction without external field as the value of the running coordination function at $d(0)$. The whole procedure is shown for $T^*=0.4$ in Figure S1. \rev{In practice, we obtain values of $d(0)$ ranging between 1.4 and 1.6 nm, which is large compared to the criterium we used in MD simulations (0.7 nm). This confirms that the geometric distinction between free ions and pairs is blurry. We want to underline that this method is only a proxy to extrapolate the free ion fraction defined through transport measurement.}

We use this method to compute the free ion fraction without an external field as a function of temperature $T^*$. Below $T^*\lesssim 0.15$, there is a large uncertainty as {the measured currents are small}. However, for those values, we observe a plateau very close to 1 in the running coordination function Figure \ref{fig:plateau}, in agreement with a fully paired system.

\begin{figure}[H]
    \centering
    \includegraphics{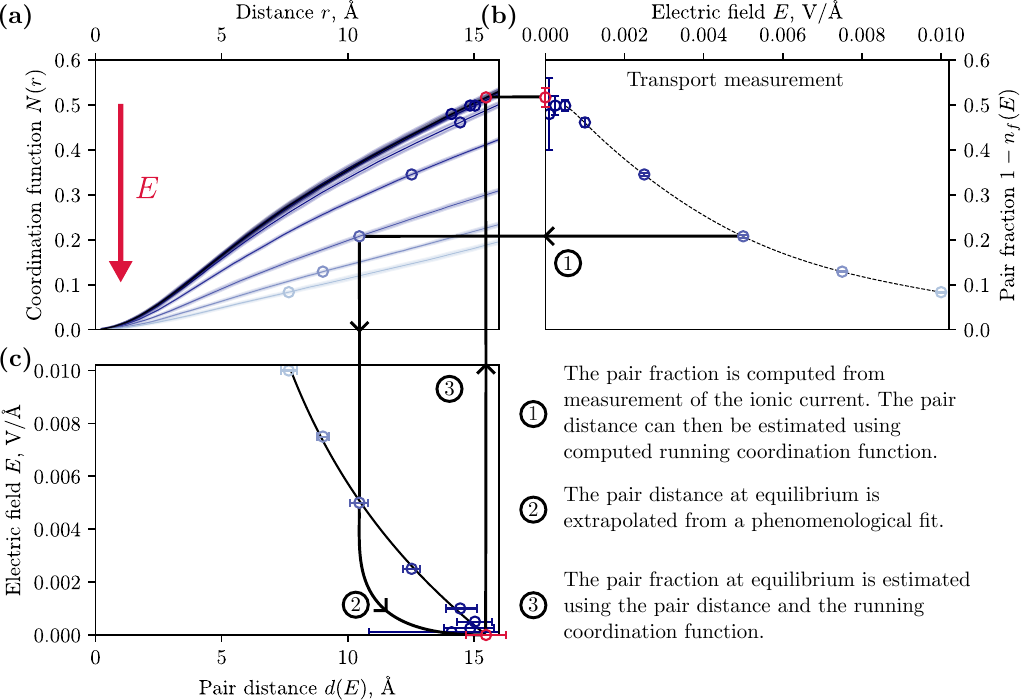}
    \caption{\label{fig:eqpairfrac} Protocol used to estimate the free ion fraction at equilibrium. (b) (Blue points) Pair fraction from NEBD simulations at $T^* = 0.4$. We compute the total ionic current from which we deduce the total free ion fraction. (a) (Curves) Running coordination function (defined in Equation 12 of main text) from NEBD (blue curves) and EBD (black curve) simulations at $T^* = 0.4$. The color of each curves match the points at a same electric field. (Blue points) Pair fraction from panel (b) are projected for each electric field. This allows us to compute the pair distance for non zero field. (c) (Blue point) Pair distance as a function of the electric field. (Dashed black curve) Fit using a phenomenological function $d(E) = d(0)\times(1+B(e^{-CE}-1))$, where $B$ et $C$ are fit parameters. This allows us to estimate the pair distance at equilibrium (Red point), with an uncertainty. Its value is reported in panel (a) to obtain the pair fraction at equilibrium. This value is also reported in panel (b) for information.}
\end{figure}

\begin{figure}[H]
    \centering
    \includegraphics{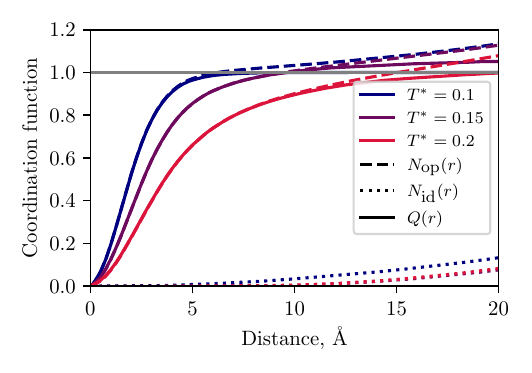}
    \caption{\label{fig:plateau} Estimation of free ion fraction at low reduced Coulomb temperature. Running coordination function of opposite ions (dashed lines), identical ions (dotted lines) and counter-charge (plain line, $N_{\text{op}}-N_{\text{id}}$) at reduced Coulomb temperature below the transition 0.1 (blue) and 0.15 (purple) and above the transition 0.2 (red). Below the transition, the running coordination function of opposite ions almost reach a plateau, at a value close to 1, at distances where the function for identical ions is very low. This yields to a quick saturation of the counter-charge close to 1: all the ions are paired. Above the transition, the counter-charge reach 1 at distances where the correlation between identical ions are non negligible, meaning that there is charge screening at larger distance than a pair. We observe that the counter-charge density goes above 1, but this seems to be a consequence of the uncertainty of the correlation functions (in particular the correlation between identical ions at small distance).}
\end{figure}

\subsection{Free charge carrier density in MD simulations}\label{sec:nf_md}

\begin{figure*}
    \centering
    \includegraphics{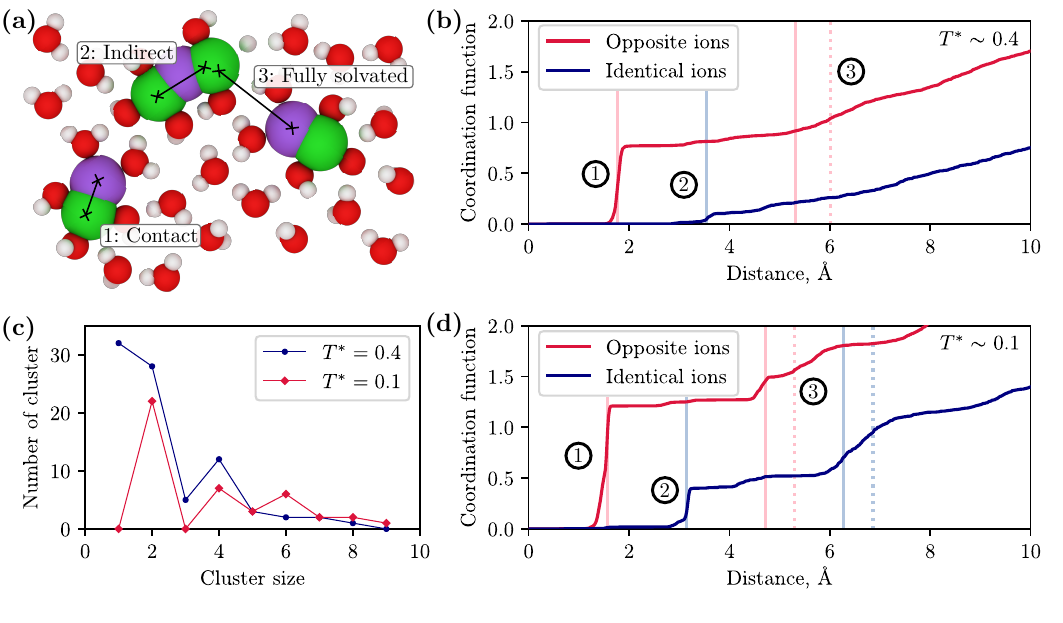}
    \caption{\label{fig:mdspike} Structure in explicit solvent simulations and its relation with correlation functions (a) Screenshot from MD simulations at equilibrium for $T^* = 0.1$. We observe 2 types of direct bonding, the contact bonds (1) when there is no water between ions, and solvent separated bonds (3) when both ions are bond, with solvent in between. We also observe indirect bonding (2) between ions, which is a consequence of the presence of clusters. (b) and (d) Running coordination function as a function of the distance between the ions, respectively for a reduced Coulomb temperature of 0.4 (monovalent ions) and 0.1 (divalent ions). The red and blue curves corresponds respectively to correlation between opposite and identical ions. We also identifies some of the well defined jump, which corresponds to spike in the correlation function. The plain pink vertical line corresponds to the contact bonds. The plain pink and light blue line corresponds to multiples of the distance between contact bond, and corresponds to indirect bond in linear chains of ions with alternating sign. The dotted pink and blue vertical line corresponds to solvent separated bonds. (c) Cluster size distribution $N_l$ defined as the number of clusters of connected ions (ions at distances smaller than $7\AAA$ from at least one other ion in the cluster). In blue and red respectively the data for a reduced Coulomb temperature of $0.4$ (monovalent ions) and $0.1$ (divalent ions).}
\end{figure*}

Figure \ref{fig:mdspike} shows respectively the oppositely and identically charged ions running coordination numbers $\nop$ and $\nid$, for monovalent ions ($T^*\sim 0.4$, panel (b)) and divalent ions ($T^*\sim 0.1$, panel (d)). Both display well defined jumps, corresponding to thin peaks in the correlation function i.e. stiff minima in the free energy. This is a direct consequence of the presence of repulsive interaction.

From comparative analysis between the simulations movies \ref{fig:mdspike}.(a) and coordination curves \ref{fig:mdspike}.(b) and (d), we can relate the structures and the peaks. We observe both direct bonds between opposite ions (non mediated by other ions) and indirect bonds (mediated by other ions, could not exist in an ionic pair). Most of the direct bonds are contact bonds at small distances $d\sim 2\ \AAA$. We also observe some solvent separated bonds, mediated by water molecules, at larger distances $d\sim 6\ \AAA$.

The position of the peaks associated to indirect bonds gives us information on the structure of the clusters. For example in our system, we mostly observe peaks at multiples of the size of contact bond, which is a signature of the predominance of linear ionic chains, in agreement with was can be seen in simulations movies.

As in Brownian dynamics we could use transport simulation in order to study pairing. Indeed, from the definition of the free charge carrier fraction:
\begin{equation}
    n_f^{\text{cluster}}(E) = \frac{j(E)}{j_0(E)},
\end{equation}
if we neglect any variation of the mobility of the field. However, as in BD simulations, this definition is not adapted to very small field. Instead, we can use the following geometric criterion.

We can define a cluster as an ensemble of ions that are at distance smaller than a distances $d_c$ to at least another ion in the cluster. $d_c$ is a small distance, below which we consider 2 opposite ions to be bound. We will use in the following $d_c = 7\AAA$ for every charge. This allows to take into account both the contact bonds (vertical pink lines in \ref{fig:mdspike}) and the solvent separated bonds (vertical dashed pink lines in \ref{fig:mdspike}).

The distribution $N_l$ of clusters of size $l$ is shown in Figure \ref{fig:mdspike}.(c) for a reduced Coulomb temperature of 0.4 (in blue), corresponding to monovalent ions, and 0.1 (in red), corresponding to divalent ions. We observe a decay of the cluster numbers with their size, and clusters with even number of ions are more abundant.

This can be explained by the fact that clusters with an odd number of ions are charged, as they cannot have the same number of anions and cations. However, we only observe neutral even clusters. The disparity is a consequence of the free energy cost to have a charged cluster. As expected, the disparity between odd and even clusters increase when the ionic charges is increased.

\begin{figure*}
    \centering
    \includegraphics{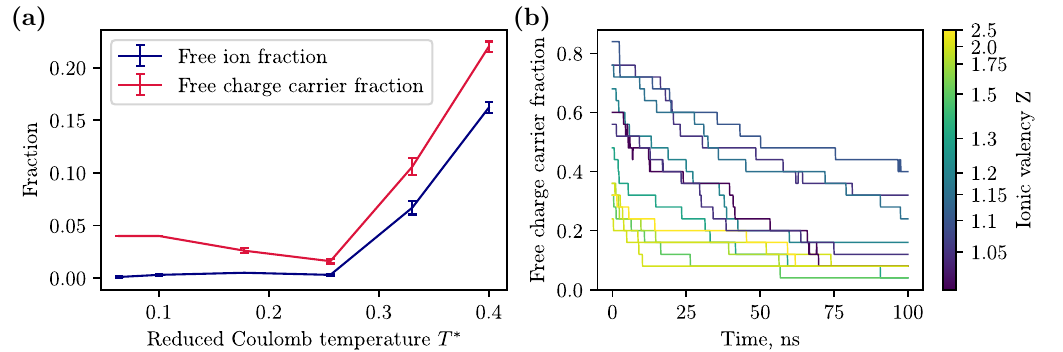}
    \caption{\label{fig:longtime} Free charge carrier density in MD simulations. (a) Free ion fraction and free charge carrier fraction as at various reduced Coulomb temperature in MD simulations. (b) Free charge carrier fraction as a function of time in MD simulations. In order to lower the simulation time, this curve is done in a system smaller ($10\times 10$nm$^2$), but with the same ionic density than the other MD simulations.}
\end{figure*}

\rev{The free charge carrier fraction and the free ion fraction are shown in figure \ref{fig:longtime}.(a). As in BD simulations, we observe a decay of both quantity at small reduced Coulomb temperature. However, the free charge carrier fraction does not decay to 0. This is because of the very slow dynamic of large cluster, which increase the time to reach equilibrium (see Figure \ref{fig:longtime}.(b)). Yet, the fraction of small charged cluster decay over shorter timescales (as shown in Figure \ref{fig:mdspike}.(c)) as small cluster have a faster dynamic. We are then able to study the transition from the density of small cluster (using the free ion fraction), despite the system still not being at equilibrium.}

\section{Non equilibrium BD simulation}
\subsection{Determination of the exponent at low field in BD simulations}
\rev{In order to study the response of the system, it is easier to compute the exponent of the association constant $n_p(E)/n_f(E)^2$ than to directly fit the conductivity. However, the presence of free ions at equilibrium is not accounted for in the theoretical model. As a consequence, in simulations, the association constant saturates at small field, and the exponent can be extracted only at large field. Instead, we define an effective association constant $K(E)$ through:
\begin{equation}
    K(E) = \frac{n_p(E)}{\delta n_f(E)^2} = \frac{1-n_f(E)}{\left(n_f(E)-n_f(0)\right)^2},
\end{equation}
that does not saturate at small field, allowing to compute the exponent from the whole exponent curve.}

\begin{figure*}
    \centering
    \includegraphics{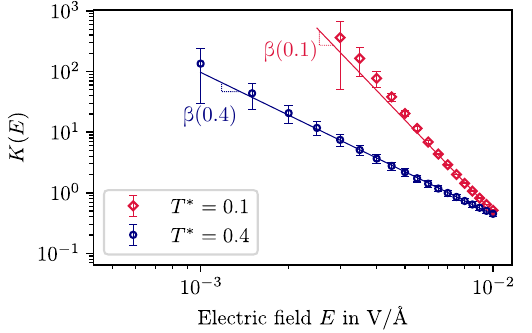}
    \caption{\label{fig:asso_cnst} Effective association constant $K(E)$ as a function of the external electric field for $T^*$ = 0.1 (red) and 0.4 (blue). (Diamond) Obtained from the BD simulations. Uncertainty is propagated from the uncertainty of $n_f(0)$ and $\delta n_f(E)$. (Line) Linear fit to extract the power law.}
\end{figure*}

In Figure \ref{fig:asso_cnst}, we plot the effective association constant as a function of the electric field. We observe that the curves are well approximated by a power law of the external field, and we can define an exponent $\beta$ obtained by fitting the curve of $K(E)$ with $E$ in logarithmic scale (Fig. 5c). At small field, we have:
\rev{\begin{equation}
    \delta j(E) \propto \delta n_f(E)E = \sqrt{\frac{1-n_f(E)}{K(E)}}E\underset{E\to 0}{\propto}\frac{E}{{\sqrt{K(|E|)}}},
\end{equation}}
hence the current also follows a power law with the external field. {The absolute value underlines the fact that the effective association constant - and then the conductivity - is symmetric under field inversion.} We can thus define the exponent $\alpha(T^*)$ of the power law:
\begin{equation}
    |\delta j(E)| \underset{E\to 0}{\propto} {|E|^{\alpha(T^*)}},
\end{equation}
{which is, as expected, antisymmetric under field inversion.}
This exponent can be extracted from the slope from \ref{fig:asso_cnst}, using $\alpha = 1 - \beta/2$.

We checked that this exponent only depends on the strength of the interaction and not other interaction parameters, such as the short-distance cut-off $C$. The exponent obtained from our BD simulations is also in agreement with the exponents obtained in previous non equilibrium MD simulations of the system \cite{robin_modeling_2021}.

\subsection{Anisotropic transport}
\begin{figure}[H]
    \centering
    \includegraphics[width=0.8\textwidth]{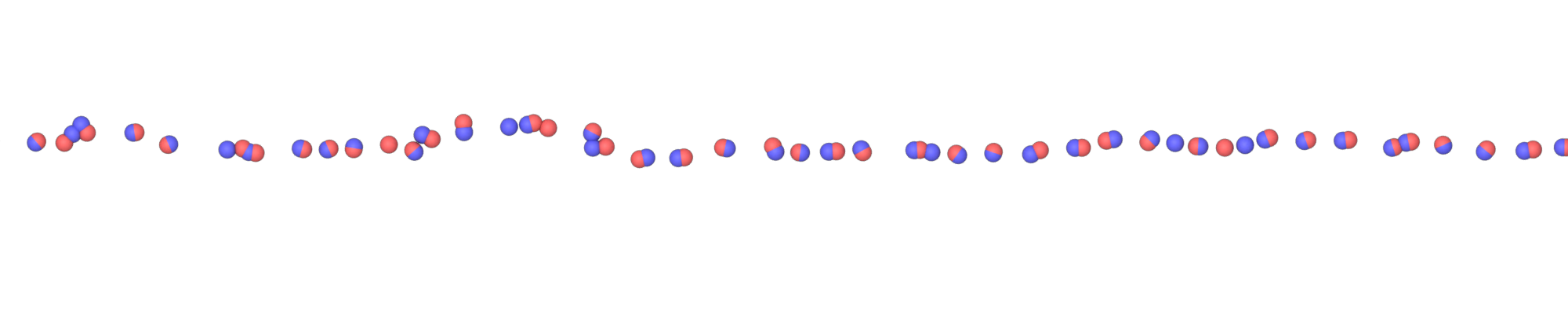}
    \caption{\label{fig:ioniclines} Screenshot of the ionic lines observed in BD simulations at low temperature $T^* = 0.01$ and large electric field $E = 0.012$ V/\AA. Despite being nonphysical - as neglecting the short distance repulsion allows ions to go through one another - this behaviour gives a good intuition of our theoretical model. The ions are free to move along the x direction, as there dynamic are fully uncorrelated by the large electric field. However, along the y direction, the ions are still fully correlated, and want to form pair - to recover local charge neutrality - creating those small chains. At lower field (and in NEMD simulations), we do not observe this behaviour, but we observe chain of pairs (or clusters). The transport go through multiple association/dissociation of those pairs (or clusters), very similarly to the Grotthus mechanism of proton transport in water. At larger reduced Coulomb temperature $T^*\gtrsim 0.25$, we do not observe those behaviour anymore, and the ions seems allowed to move in both directions.}
\end{figure}
\bibliographystyle{ieeetr}
\bibliography{ms.bib}